\documentclass[oneside,10pt,a4paper,onecolumn]{article}

\usepackage{url}
\usepackage[T1]{fontenc}
\usepackage[dvips]{graphicx}

\begin{document}

\makeatletter
\def\@biblabel#1{[#1]}
\makeatother

\markboth{S. Banisch, T. Ara\'{u}jo, J. Lou\c{c}\~{a}.}{Opinion Dynamics and Communication Networks}

\begin{titlepage}
\title{\vspace{-2.0cm}\bf{Opinion Dynamics and Communication Networks}}
\author{Sven Banisch$^{1,2,3}$, Tanya Ara\'{u}jo$^{1,3}$ and Jorge Lou\c{c}\~{a}$^{1,4}$
                \\ \\
        \small $^1$ Institute for Complexity Science (ICC), 1649-026 Lisbon (PORTUGAL)\\
        \small $^2$Mathematical Physics at Physics Faculty, Bielefeld University, \\
        \small D-33501 Bielefeld (GERMANY)\\
        \small $^3$Research Unit on Complexity in Economics (UECE), ISEG, \\
        \small TULisbon, 1200-781 Lisbon (PORTUGAL)\\
        \small $^4$ Laboratory of Agent Modelling (LabMAg), \\
        \small ISCTE, 1649-026 Lisbon (PORTUGAL)\\
        \small \tt [sven.banisch@universecity.de | tanya@iseg.utl.pt | Jorge.L@iscte.pt]\\\vspace{0.0cm}
        }
\date{}
\end{titlepage}

\maketitle

\begin{abstract}
This paper examines the interplay of opinion exchange dynamics and communication network formation.
An opinion formation procedure is introduced which is based on an abstract representation of opinions as $k$--dimensional bit--strings. Individuals interact if the difference in the opinion strings is below a defined similarity threshold $d_I$. Depending on $d_I$, different behaviour of the population is observed: low values result in a state of highly fragmented opinions and higher values yield consensus.
The first contribution of this research is to identify the values of parameters $d_I$ and $k$, such that the transition between fragmented opinions and homogeneity takes place.
Then, we look at this transition from two perspectives: first by studying the group size distribution and second by analysing the communication network that is formed by the interactions that take place during the simulation.
The emerging networks are classified by statistical means and we find that non--trivial social structures emerge from simple rules for individual communication.
Generating networks allows to compare model outcomes with real--world communication patterns.
\end{abstract}

%\keywords{Opinion Dynamics; Social Networks; Coevolution; Computational Models; Artificial Societies.}

\section*{Introduction}

Many societal processes are ultimately based on the mutual interactions among individuals with diverse opinions, attitudes and lifestyles.
The processes of inter--personal communication and opinion exchange play a crucial role in the formation of social structures and networks.
Examining the interplay of opinion exchange and communication network formation is the main issue addressed by this study.

Therefore, an opinion formation model inspired by the abstract agent model presented in~\cite{Araujo2009} is introduced.
Opinions are represented as a series of $k$ bits, which we find an interesting approach to the modelling of attitudes and beliefs, since human thinking can be represented in terms of polarities (yes/no, good/bad, young/old, etc.).
And moreover, we are used to measuring information in bits.

Such an abstract bit--string approach has been used in the
simulation of consumer--producer behaviour~\cite{Araujo2009} as
well as in the context of labour market analysis~\cite{Araujo2008}
where bit--strings represent products (or job offers) and needs
(worker skills). Here, each bit--string represents an agent opinion
and a procedure of agent--agent interaction is specified based on
assumptions from social comparison theory~\cite{Festinger1954} and
in opinion formation
models~\cite{Axelrod1997,Deffuant2000,Amblard2004,Schweitzer2003,Schweitzer2008}.

This paper is organized in the following way. We start reviewing
previous approaches to the modelling of opinion exchange dynamics.
After this, we give an explanation of our model.
This is followed by a
numerical analysis, in which the opinion evolution is considered
before looking at the emerging networks of communication activity.
A discussion of the results concludes this work.

\section*{Related Work}
\label{sec:2}

The most important observation from studying computer models of social influence and opinion dynamics is probably that interaction rules by which interacting agents tend to become more alike in their beliefs do not necessarily lead to a population in which all the individuals share the same opinion.
To put it in Axelrod's words (\cite{Axelrod1997}, p.223): >>Local convergence can lead to global polarization.<<
This dynamic effect which contrasts common intuition has been shown and analysed in a large number of different opinion formation models (e.g.,~\cite{Axelrod1997,Castellano2000,Deffuant2000,Schweitzer2003,Amblard2004,Holme2006}).

All these models are based on two principles: (1.) two individuals are more likely to communicate with one another if they already share a number of opinion features (i.e., their opinions are similar); and (2.) communication further increases this number of shared features (i.e., individuals become even more alike).
Approaches differ mainly in their representation of opinion.
Some models use continuous representations (e.g.,~\cite{Deffuant2000,Amblard2004}) whereas others assume opinions to be a set of features which can take different (discrete) values~\cite{Axelrod1997,Castellano2000}.
Also populations in which agents may have only two possible choices (yes/no) have been studied frequently (e.g.,~\cite{Schweitzer2003,Galam2004,SznajdWeron2004,Zanette2006,Jiang2007}).

Depending on the control parameters, opinion models give rise to
quite a variety of population structures, from a highly fragmented
population in which only few individuals share the same opinion, to
homogeneity or, in between these two regimes, to a stable state with
several differently sized groups. These three \emph{behaviour
classes}\footnote{Behaviour classes refer to qualitative different
behaviours that potentially result from a simulation model.} have
been reported in several previous studies on opinion dynamics
(e.g.,~\cite{Axelrod1997,Castellano2000,Deffuant2000,Schweitzer2003,Amblard2004,Holme2006,Zanette2006}).
Moreover, in the transition regime, group sizes were found to follow a power law distribution~\cite{Castellano2000}.
For a detailed report with particular focus on such statistical properties of social dynamics processes the reader is referred to~\cite{Castellano2009}.

The social network structure which is assumed to underlie the
opinion exchange dynamics is another important aspect. In opinion
formation studies as well as in other disciplines, often a static
network determining which agents may interact with one another is
imposed. In social influence models this is referred to as an
agent's neighbourhood. Only recently, the co--evolution of agent
states and networks formed by processes (based on the states)
receive more attention (see~\cite{Gross2008} for an
inter--disciplinary, recent review). Simultaneously, adaptive
network approaches that focus on the interplay of opinion dynamics
and network formation have been further
developed~\cite{Holme2006,Zanette2006,Rosvall2007,Amblard2008a}.

In reference ~\cite{Holme2006}, a parameter is used to determine
whether agents form their opinion based on the opinions of connected
agents or if they re--link to an agent having the same opinion. This
simple model was found to undergo the same phase transition from
diverse to homogeneous opinions as reported above. The effect of
giving agents the possibility of cutting links, if they do not
achieve agreement with their neighbours, is also studied
in~\cite{Zanette2006}. The study presented in~\cite{Amblard2008a}
starts from a random network determining the possible communication
links between the agents. The similarity between two linked agents
(which evolves in time since the opinions of interacting agents are
updated) is used to assign a frequency of interaction. In this way,
dynamic network structures emerge without the need for specifying
conditions for cutting social links.

In this paper, we also avoid introducing further assumptions for
re--wiring the network. Instead, the network design is based
on the communication that effectively takes place among the agents.

\section*{Model Definition}

\begin{figure}[ht]
    \centering
        \vspace{-6pt}
        \includegraphics[width=.80\linewidth]{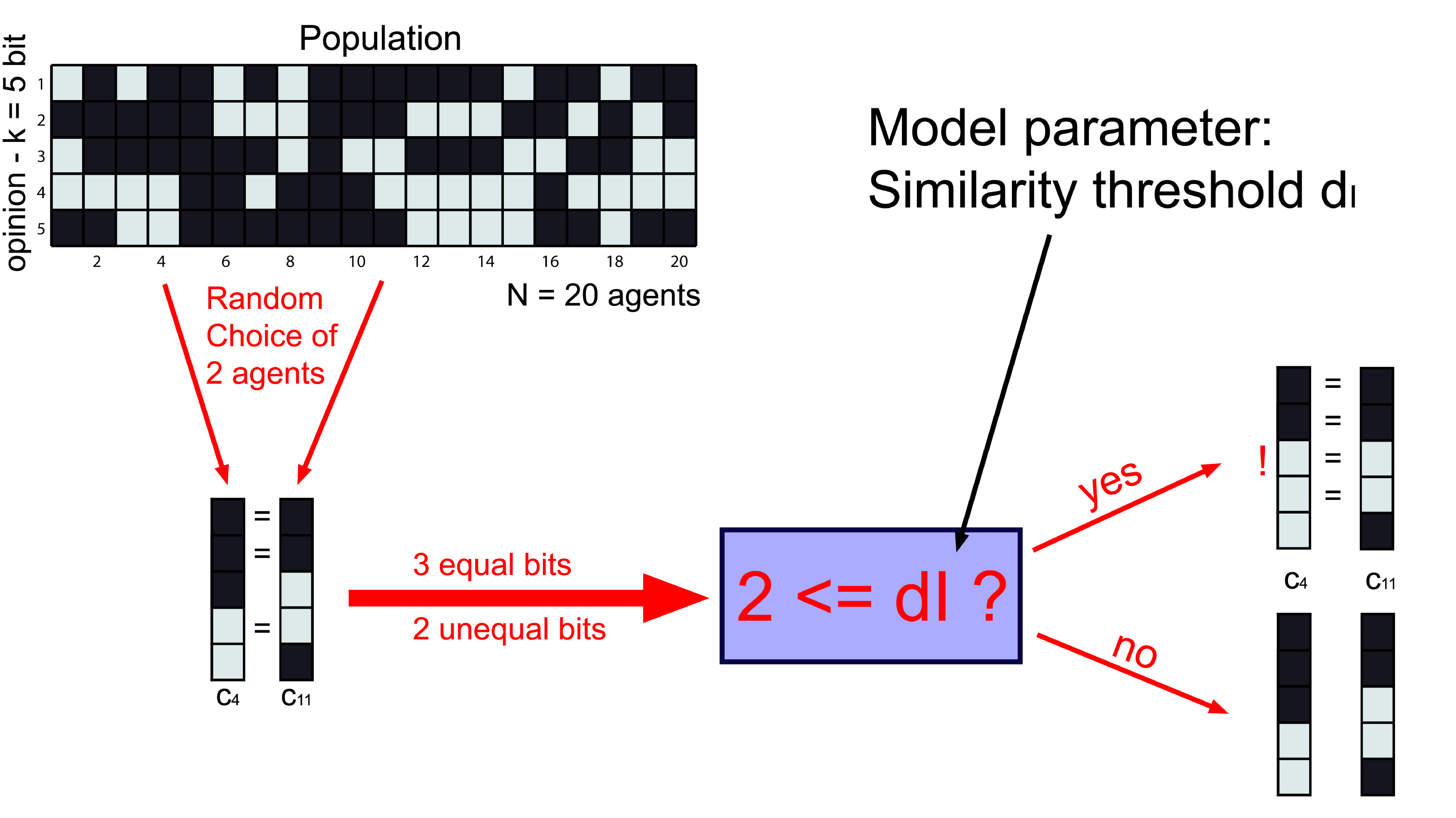}
       \vspace{-5pt}
    \caption{Illustration of the interaction process.}
    \label{fig:model}
\end{figure}

The opinion formation model implemented for this study bases on a bit--string representation as used in~\cite{Araujo2009} concerning the simulation of consumer--producer behaviour and in~\cite{Araujo2008} for the analysis of labour market dynamics.
In these examples, bit--strings represent products (or, respectively, work offers) and needs (skills).
The exchange is based on the matching of these two strings.
Here, this concept is used in the modelling of opinion exchange where a series of $k$ bits represents an agent opinion, and interaction between agents takes place if their opinion strings are similar.
Two agents are willing to interact with one another if the matching of their opinion strings is below (or equal to) a certain similarity threshold denoted by $d_I$.
The interaction process is illustrated in Fig.~\ref{fig:model}.

In the beginning $N$ agents are generated and a random bit--string
is assigned to them. In the interaction process, two agents, say
$c_4, c_{11}$ according to Fig.~\ref{fig:model}, meet at random. But they are only willing to
communicate about an issue (one element of the bit--string), if the
number of unequal bits (i.e., the hamming distance $h$) is below or
equal to a similarity threshold $d_I$. If the hamming distance
between the two agents is below the threshold ($h(c_4,c_{11}) \leq d_I$), they exchange ideas
about an issue. As a result of that the agent chosen first ($c_4$ in the example) adopts the opinion of the
$c_{11}$ concerning that issue by flipping the respective
bit.\footnote{Since the probability of an agent to be the first one
is equal for all individuals during the iteration process, all the
agents have an equal chance to imitate or to be imitated. The same
assumption about who imitates who is made by Axelrod in~\cite{Axelrod1997}.} In this way
the principles of similarity and imitation which form the basis of
most opinion models
(e.g.,~\cite{Axelrod1997,Castellano2000,Deffuant2000,Schweitzer2003,Amblard2004,Holme2006})
are integrated.

The model is implemented so that during a single time step all the agents have the chance to interact with one another.
Therefore one time step of the model corresponds to $N$ interactions of pairs of randomly chosen agents $c_i,c_j$.
We exclude self--interaction ($c_i = c_i$), but we do not force all agents to be chosen exactly one time (with the result that some may be chosen twice and other ones are not chosen at all in that iteration).

In order to understand better the opinion exchange dynamics among the agents, we keep track of the interactions that take place in the course of the simulation.
We introduce an interaction matrix $I$ which stores all the interaction activity.
$I$ is of size $N \times N$ and the element $i_{ij}$ saves the number of times $c_i$ and $c_j$ interacted in some way.
Therefore, each time two chosen agents $c_i,c_j$ are sufficiently similar ($h(c_i,c_j) \leq d_I$), we increase $i_{ij}$ and $i_{ji}$ by one.
Note that $i_{ij}$ and $i_{ji}$ are also increased if agents already share the same opinion.
The matrix $I$ corresponds to a weighted graph, in which edges represent the communication lines between different agents.
Additional information about who imitates who, and which agents are imitated is stored separately.

To summarize. Starting from an initial random population, at each
time step $N$ pairs of agents ($c_i$, $c_j$) are chosen, and if the
distance is below the threshold ($h(c_i,c_j) \leq d_I$), the agent $c_i$ switches one of the bits that
have been unequal. In other words, if individuals are close enough
they have the opportunity to become even closer. Otherwise, they
just do not communicate at all. As they split apart groups are
formed. In the next section, we discuss the characteristics of the
groups that emerge from this process.

\section*{Behaviour Classification}

In order to obtain a classification
of the general dynamic behaviour of the model, a series of
systematic tests has been performed. For this purpose, we looked at
different numbers of dimensions $k$ used in the opinion
representation and respectively different threshold values $d_I$.
Note that considering the ratio of the two parameters
$\frac{d_I}{k}$ might also be of interest, since it accounts for the
relative similarity required for two agents to interact. It would
also be favourable as the number of model parameters would be
reduced.

However, in order to be clear about the interdependence of the two
parameters, we first tested all the configurations $k = 1 \ldots 32$
and respectively $d_I = 1 \ldots k$, and looked at the number of
groups of individuals that share exactly the same opinion (denoted
by $N_{G}$). A group in this sense can be formalized as
\begin{equation}
G_o = \left\{ c_i : h(c_i,o) = 0 \right\},
\label{eq:Go}
\end{equation}
where $o$ is the reference opinion string shared by all the members $c_i$ of the group $G_o$.
A particular group $G_o$ comprises all the agents $c_i$, whose opinions equals $o$,   (that is the hamming distance $h(c_i,o) = 0$).
Consequently, $N_G$ is defined as the number of groups $G_o$ with at least one member.
With groups defined in this way, the maximum number of possible groups is
\begin{equation}
\max(N_G) = \min(N,2^k).
\label{eq:N_G}
\end{equation}
As opinions are represented as a series of $k$ bits, there are $2^k$ possible opinion strings.
But in the case that the number of agents is below that number ($N < 2^k$), then the maximum number of groups with at least one member is equal to $N$ (i.e., $\max(N_{G}) = N$).

The model behaviour can be classified by using $N_{G}$ as an
indicator for different kinds of behaviour. The case where $N_G$ is
near the theoretical maximum ($N_G \approx N$ or respectively $N_G
\approx 2^k$ as described above) represents the cases in which the
public opinion remains highly fragmented, since there are many
groups with only few members (or even just a single one). The other
extreme is represented by $N_G = 1$ in which case all the agents
belong to a single giant group, that is: the society of agents reaches global consensus.

\begin{figure*}[ht]
    \centering
    \vspace{-11pt}
        \includegraphics[width=1.0\textwidth]{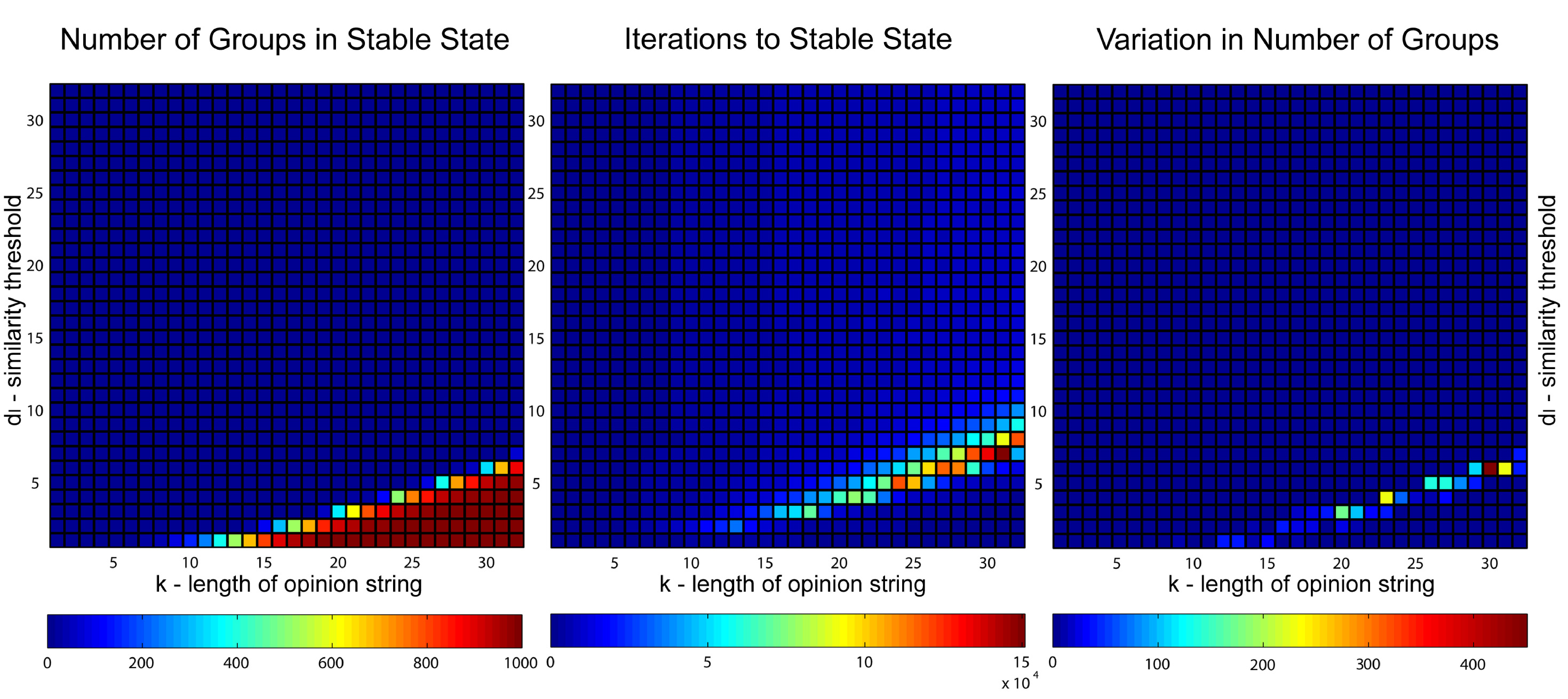}
    \vspace{-33pt}
    \caption{On the l.h.s., classification of the model behaviour in terms of $N_{G}$ with respect to $k$ and $d_I$ for 1000 agents. The middle image shows the iteration number required to reach the stable state, and on the r.h.s. the variation of $N_{G}$ is shown. Average of 5 simulation runs.}
    \label{fig:BehaviourClassificationKdI}
\end{figure*}

In Fig.~\ref{fig:BehaviourClassificationKdI} this classification of the model behaviour with respect to $k$ and $d_I$ is shown using a $32 \times 32$ parameter grid formed by $k = 1 \ldots 32$ and $d_I = 1 \ldots k$.
$N = 1000$ agents have been used for the experiments and five simulation runs have been performed for each parameter configuration.\footnote{On the whole, $5 \times 528$ runs with $N = 1000$ agents have been performed.}
The two additional grids represent the iteration number required to reach the stable state (middle) and the variance of $N_{G}$ (r.h.s) observed over the five realizations.
Note that the stable state, in which no further opinion exchange is possible, is always reached, though the number of iterations required to reach it differs tremendously.

Fig.~\ref{fig:BehaviourClassificationKdI} makes clear that a transition takes place from a population in which all opinions are the same (blue region with $N_G \approx 1$) to a population in which basically all the agents have different opinions (dark red with $N_G \approx 1000$).
The third and in fact most interesting behaviour is observed in the area of transition in between these two extremes.
It will be considered using a specific parameter constellation in the following section.

%\enlargethispage{\baselineskip}
\section*{Group Size Distribution}

\begin{figure}[ht]
    \centering
        \includegraphics[width=.90\linewidth]{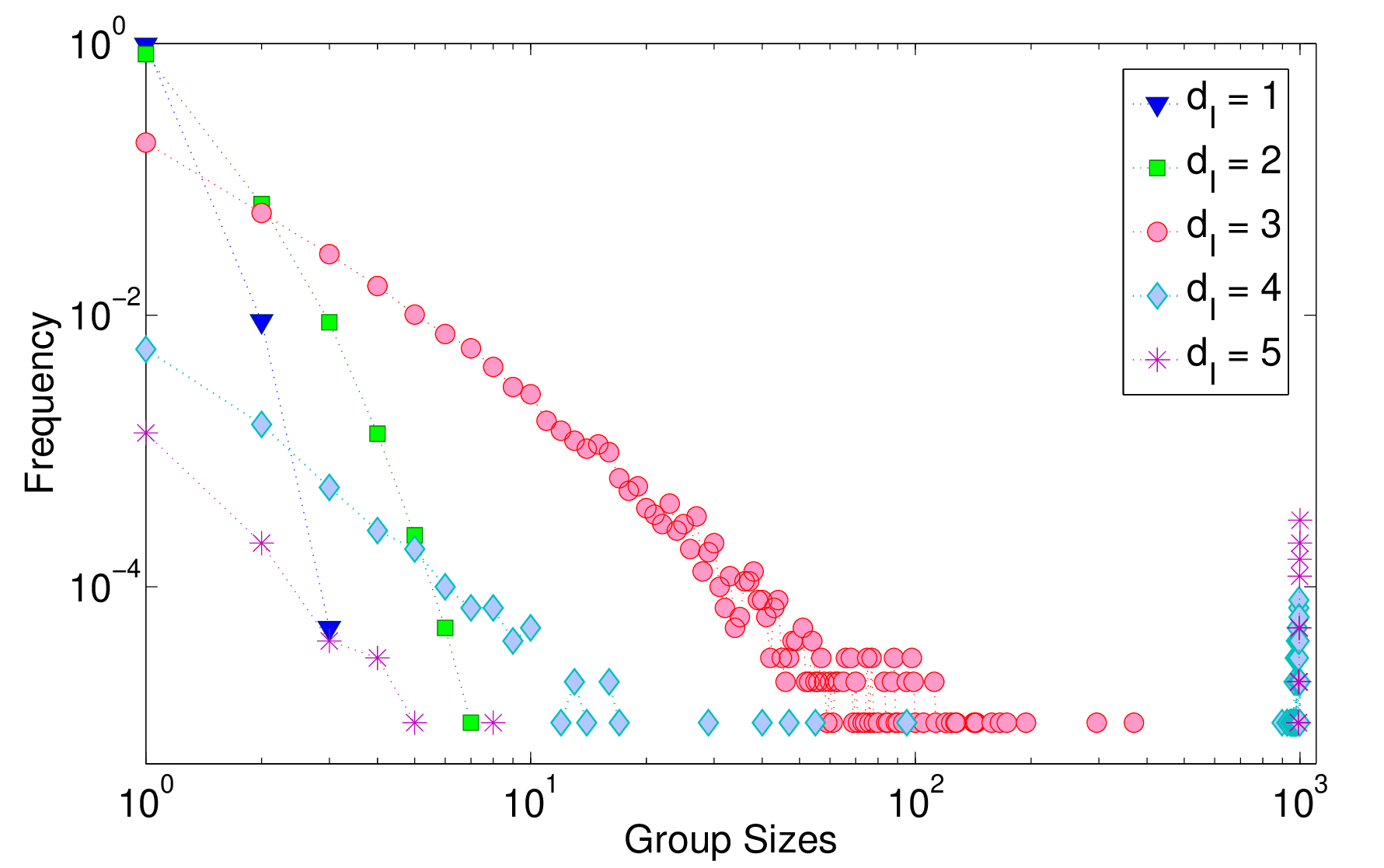}
    \caption{Group size distribution in stable state for $k = 20$ and $d_I = 1\ldots5$ based on 100 simulation runs.}
    \label{fig:GroupSizes}
\end{figure}

In order to get a better idea of the model behaviour in transition, we concentrate on the example $N = 1000$ and $k = 20$ in the following two sections.
The influence of the population size $N$ is studied after that.
From the images in Fig.~\ref{fig:BehaviourClassificationKdI} we assume that the critical behaviour can be observed for $d_I = 3$, where an average number of groups $N_G \approx 400$ was found with a very high variance.
In a second series of systematic experiments, now using 100 simulation runs, we subsequently increase the threshold from $d_I = 1 \ldots 5$ and look at the distribution of group sizes in stable state.
The results are presented in Fig.~\ref{fig:GroupSizes}.

It becomes clear that, for $N=1000$ and $k=20$, $d_I = 3$ is indeed
at the border between the two qualitatively different behaviours:
homogeneity and fragmentation. It therefore displays the model
behaviour in the phase transition. Moreover,
Fig.~\ref{fig:GroupSizes} shows that groups scale according to a
power law for $d_I = 3$, which was also reported in the phase transition of the Axelrod model in~\cite{Castellano2000}.

For $d_I < 3$ only very small groups are present, whereas for $d_I > 3$ the likeliness of small groups to form subsequently decreases and agents are very likely to meet in a single giant group (global consensus).
Note, however, that for $d_I = 4$ several intermediate group sizes are also observed that show a power law scaling for group sizes up to 20 members.

\section*{Communication Networks}

The analysis of the groups that consist of individuals sharing the same opinion is an interesting issue, looking at the communication activities that led to this state, on the other hand, can reveal important additional information of how a certain state is reached.
From the network point of view, we can consider that agents are the nodes of a communication network and that edges represent communication lines.
In order to keep track of the communication activity within such a network, we introduced the interaction matrix $I$ which stores all the interactions that take place in the course of the simulation.
We also compute the adjacency matrix $A$, the elements of which are
$a_{ij} = 0$ if $i_{ij} = 0$ and $a_{ij} = 1$ if $i_{ij} > 0$.
Matrix $A$ accounts for communication lines between the agents, but
not for the intensity (i.e., the frequency) of communication between
them.

\begin{figure}[ht]
    \centering
        \includegraphics[width=.90\linewidth]{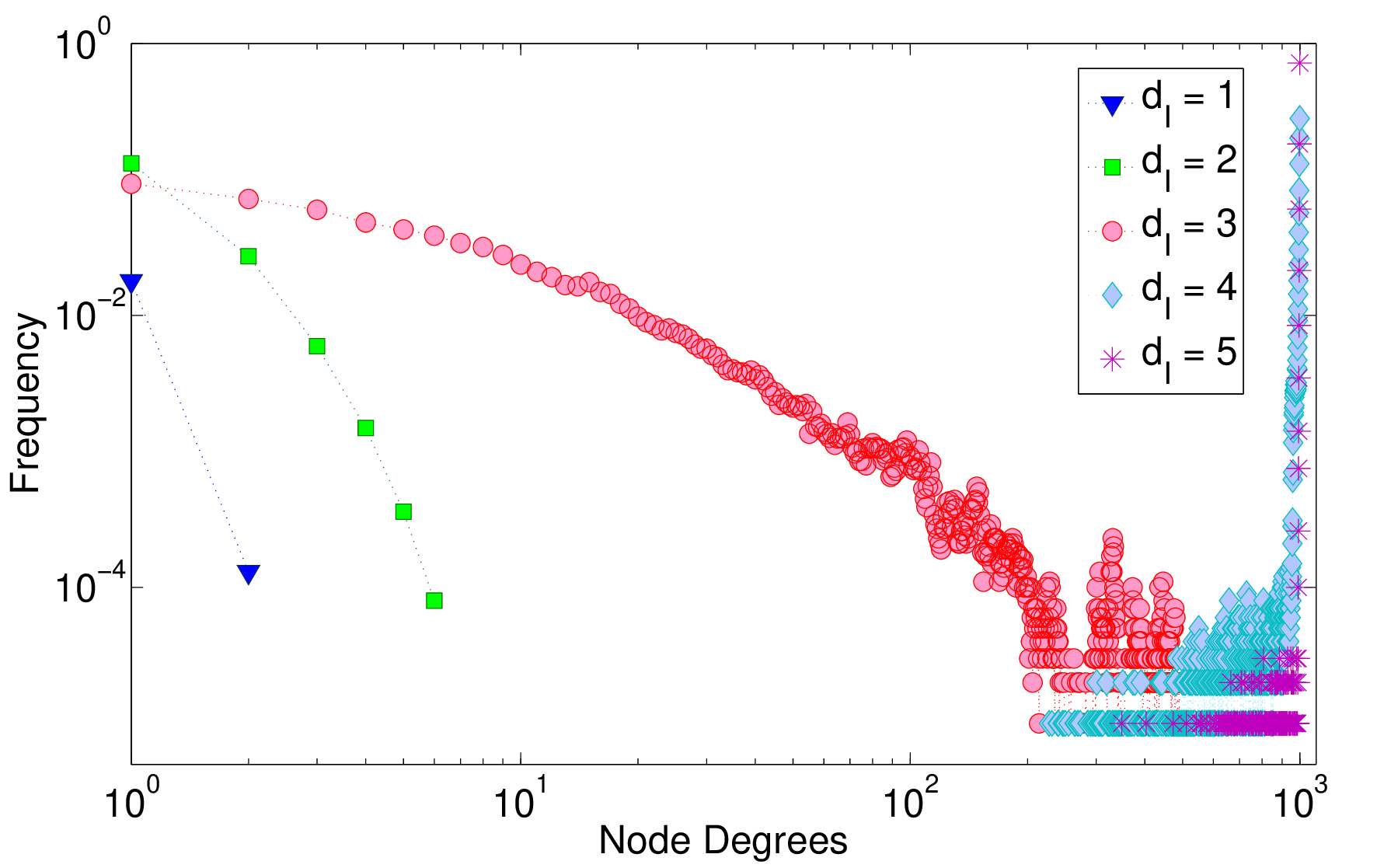}
    \caption{Log--log plot of the degree distribution for $N = 1000$, $k = 20$, and different $d_I = 1 \ldots 5$. The distribution for $d_I = 3$ displays the behaviour in the phase transition.}
    \label{fig:DegreeDistribution}
\end{figure}

In Fig.~\ref{fig:DegreeDistribution} the degree distribution of the communication network computed on the basis of $A$ is visualised.
Degree distributions are used frequently in the classification of complex networks (e.g.,~\cite{Albert2002,Barrat2004,Gross2008,desJardins2008}).
The data displayed in Fig.~\ref{fig:DegreeDistribution} represents the same 100 realisations performed for the group size analysis shown in Fig.~\ref{fig:GroupSizes}.

%\enlargethispage{\baselineskip}
%\enlargethispage{\baselineskip}

The distribution shows that for $d_I = 1$ the
frequency with which node degrees of one or two (represented by the
blue triangles) have been observed is around $0.02$, which makes
clear that the great majority of the nodes has a zero degree, i.e.,
they are isolated nodes. For $d_I = 2$, still, the majority of nodes
is isolated, but there are agents connected to up to six others.
All in all, for low values of $d_I$ the differences between the agents maintain, because initially, the distance between most agents is larger than $d_I$.
Therefore, interaction becomes unlikely and the communication network is very weakly connected.

The degree distributions for $d_I = 4$ and $d_I = 5$ are quite different from the prior examples.
They display the characteristics of a highly connected (quasi--complete) network.
In the course of 100 simulations no node with a degree below 200 was found which means that the weakest connected agent is still connected to more than 200 other agents.
Moreover, the larger $d_I$ is chosen, the more close the final network is to the complete graph.\footnote{For $d_I = 6$ the percentage of node degrees smaller than 900 was 0.08\%.}
This is because potentially all the agents are allowed to interact with all the others (complete graph), and with an increased threshold value more of these interactions really take place since the condition $h(c_i,c_j) \leq d_I$ is more likely to be satisfied.

An interesting property is observed when comparing the degree distribution for $d_I = 4, d_I = 5$ to the group size distribution in Fig.~\ref{fig:GroupSizes} as it results using the same values $d_I$.
As shown in Fig.~\ref{fig:GroupSizes}, small minority groups of agents may form in these cases, although these agents interacted with at least 200 others (as we know from Fig.~\ref{fig:DegreeDistribution}), i.e., they are not isolated at all.
We conclude that intensive communication behaviour does not automatically make agents adopt the opinion of the largest group, agents (or small groups of agents with up to ten members) may have an individual >>outsider<< opinion even though the communication activity involved many agents from different groups.

From the network point of view the phase transition from highly fragmented opinions to global consensus as reported above now becomes the transition from a weakly connected network (i.e., reduced communication activity for $d_I = 1,2$) to a complete network (in which all the agents communicate with all the others for $d_I = 4,5$).
The degree scaling behaviour in between these two regimes ($d_I = 3$) represents the model behaviour in the phase transition.

The distribution of the degrees for $d_I = 3$ as shown in Fig.~\ref{fig:DegreeDistribution} displays the characteristics of a non--trivial network structure (compare also the network shown in Fig.~\ref{fig:Network1000}).
The distribution is significantly different from distributions found in random networks.
Fig.~\ref{fig:DegreeDistribution} indicates that the scaling of the degrees is according to a power law, at least for the upper tail with degrees below 100.
This means that the network is scale--free.
%, i.e., the nodes connectedness does not vary with the scale of the network.
For connectivities around 100 slight deviations are visible and for degrees in the region from 300 to 500 we cannot any longer assume a power law behaviour.

Further analysis using a larger number of experiments\footnote{In the physics literature (e.g.,~\cite{Castellano2000,Holme2006}) often $10^4$ realizations are used.} might reveal the reasons for these irregularities, and they will also make us more confident about the power law distribution indicated by Fig.~\ref{fig:DegreeDistribution}.
Another equally important issue to provide us with a more complete picture of the model behaviour is the influence of the populations size.

\section*{Influence of the Population Size}

In order to be able to compare the model outcomes for realizations using different numbers of agents ($N$), it is convenient to use the relative number of groups after stabilization given by $\frac{N_G}{N}$.
In the case of opinion fragmentation $\frac{N_G}{N} \approx 1$ since $N_G \approx N$.
In the case a global consensus is reached we have $N_G = 1$ and therefore $\frac{N_G}{N} = \frac{1}{N} \approx 0$.
For the intermediate states the relative number of groups stays between zero and one ($0 < \frac{N_G}{N} < 1$).
\begin{figure}[ht]
    \centering
    \vspace{-11pt}
        \includegraphics[width=1.0\linewidth]{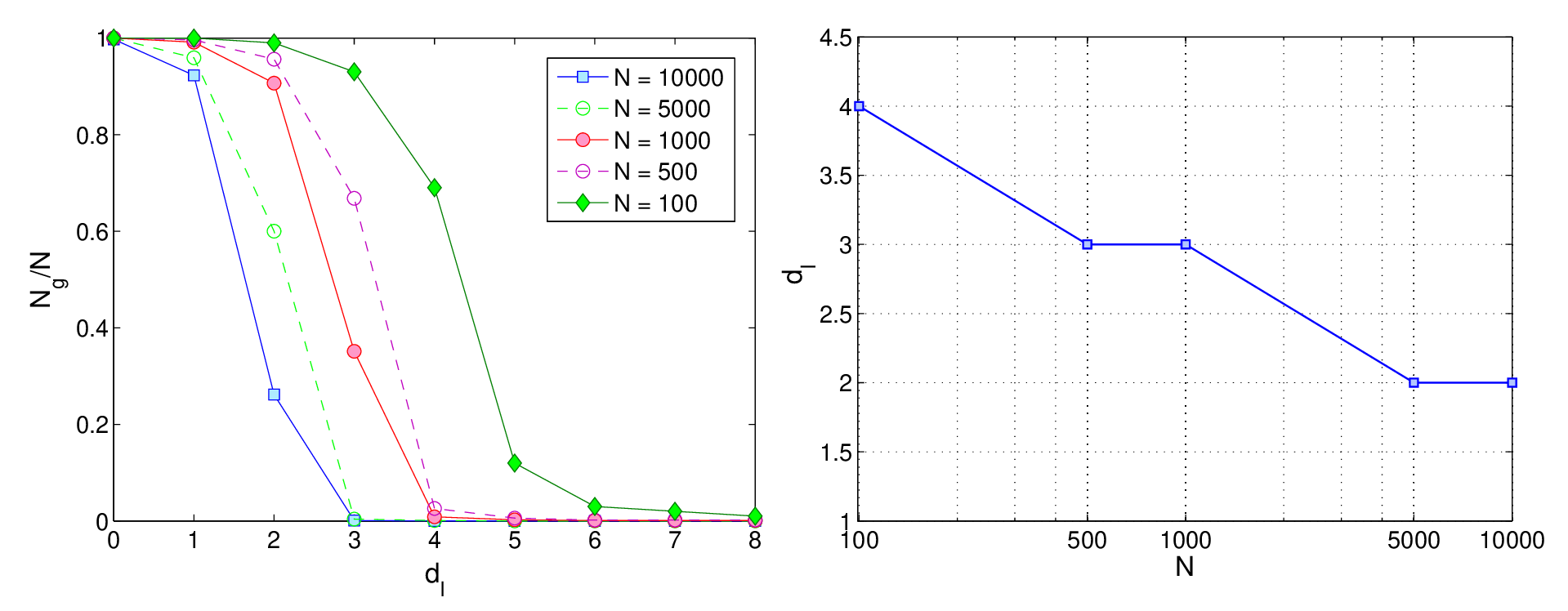}
     %\vspace{-22pt}
    \caption{Dependence of the model behaviour on the population size $N$. L.h.s.: The relative number of groups after stabilisation ($\frac{N_G}{N}$) is shown for different population sizes $N$. R.h.s.: The critical values $d_I^*$ at which transition behaviour is observed with respect to the population size. Results for 10 realizations and $k = 20$.}
    \label{fig:NgToN}
\end{figure}

In the l.h.s. of Fig.~\ref{fig:NgToN}, the relative number of groups in stable state is shown for $N=100,500,1000,5000$ and $10000$ as a function of the threshold $d_I$. 
In all simulations an opinion string of length $k = 20$ is used.
Each data point in Fig.~\ref{fig:NgToN} represents the average of $\frac{N_G}{N}$ determined over 10 repeated simulation runs.
Let's call $d_I^*$ the threshold values at which the transition from fragmentation  
to consensus takes place. 
With each increase of the population size from $N = 100$ to $N = 10000$ the values at which transition behaviour is observed ($d_I^*$) decreases.

The r.h.s of Fig.~\ref{fig:NgToN} makes this clear.
It shows the >>critical<< values $d_I^*$ as a function of the population size.
For a certain $N$, $d_I^*$ has been chosen to correspond to those 10 repeated simulation runs in which the variance in the values of $\frac{N_G}{N}$ is maximal, as a high variance is a suitable indicator for transition behaviour (compare also Fig.~\ref{fig:BehaviourClassificationKdI}).
The r.h.s of Fig.~\ref{fig:NgToN} shows the decrease of the $d_I^*$ with the population size $N$.
Note the logarithmic scale used for the number of agents $N$ on the horizontal axis.
As the $N$ grows 10 times larger, the critical threshold decreases by one.

\begin{figure}[ht]
    \centering
        \includegraphics[width=.9\linewidth]{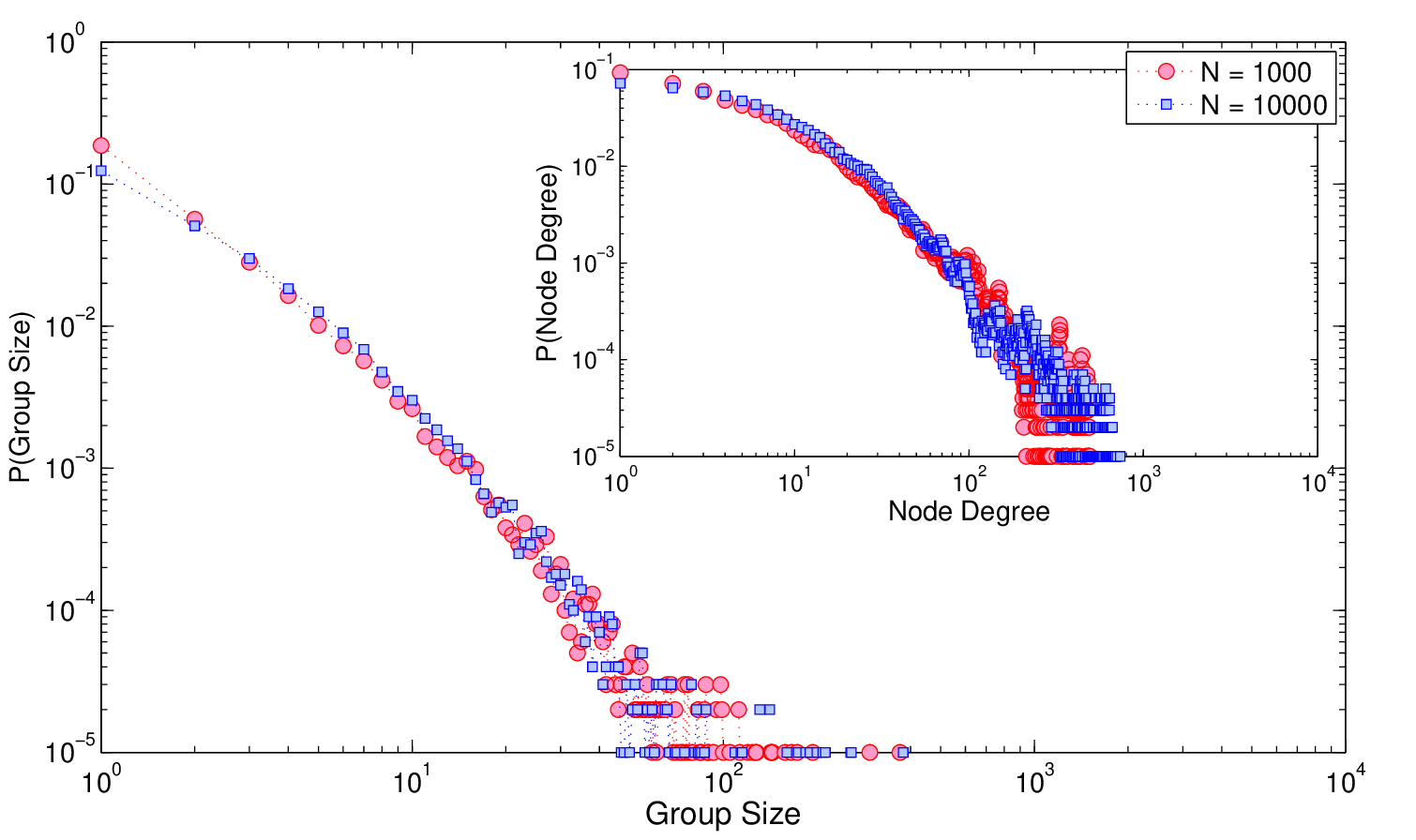}
    \caption{Group size and degree (inset) distributions for $N = 1000, d_I = 3$ and $N = 10000, d_I = 2$ with $k = 20$ in both cases. The distributions for N=1000 and N=10000 are based on, respectively, 100 and 10 repeated simulation runs.}
    \label{fig:All.N1000to10000}
\end{figure}

In Fig.~\ref{fig:NgToN} the influence of $N$ on the model behaviour is evaluated on the basis of the average value and the variance of a series of relative numbers of groups, $\frac{N_G}{N}$.
A further interesting aspect is the influence of the population size on the group sizes in stable state and the interaction network formed as the result of the opinion exchange process.
For $N = 1000, d_I = 3$ and $N = 10000, d_I = 2$ (both displaying the characteristics of transition behaviour) the group size and the degree distributions are presented and compared in Fig.~\ref{fig:All.N1000to10000}.

We observe in Fig.~\ref{fig:All.N1000to10000} that there is no remarkable difference between the groups formed in a population of 10000 agents compared to the example of 1000 agents considered previously.
The degree distribution shown in the inset of Fig.~\ref{fig:All.N1000to10000} displays the same scaling behaviour for both cases as well.
This indicates that in the phase transition the observed interaction patterns as well as the statistical properties of the grouping behaviour after stabilisation is invariant to changes of the population size.
Moreover, observing the same power law scaling of the degrees for $N = 1000$ and $N = 10000$ confirms that the opinion exchange process as presented in this paper yields scale--free communication networks.

The observed power law scaling can be an interesting issue for the validity of the model, since such a behaviour is shown to be a property of many real--world networks (compare~\cite{Albert2002,Barrat2004,Chung2006,Onnela2007a,Onnela2007b,Leskovec2008}). 
The invariance of the communication patterns to the population size $N$ (as indicated by Fig.~\ref{fig:All.N1000to10000}) is another important issue for the comparison with real--world network data, since empirical network studies often consider a large number of participants (up to several millions).

\section*{Networks as a Link to Reality}

Recently, a critical account by Sobkowicz (\cite{Sobkowicz2009}) revealed that among the vast literature on opinion formation models, there are very few attempts to link the simulation results to real--world data.
The main reason for this is that high--quality data of opinion spread in real societies is not available.
However, recent analyses of real--world communication networks (\cite{Onnela2007a,Onnela2007b,Leskovec2008}) have led to a better understanding of human communication activity.
The network view on opinion exchange processes introduced in this paper, enables that simulation outcomes be compared to this new insights.
This may be of great value in the refinement and calibration of opinion models.

One example of networks that can be adequate in the comparison are phone call networks (e.g., \cite{Aiello2000,Onnela2007a,Onnela2007b}).
As in our model, these networks are constructed by one--to--one communication.
In~\cite{Onnela2007a} and~\cite{Onnela2007b}, Onnela et al. extensively analyse the structure of a mobile phone network constructed by observing the communication activity of several million users during 18 weeks.
They explain that the data serves as >>a proxy for the underlying social network<< (\cite{Onnela2007a}, p. 3) and therefore of human communication patterns.
For any opinion model to be of explanatory value (as an explanatory candidate in the words of Epstein~\cite{Epstein2006}), it is necessary that realistic network structures (macro--behaviour) are formed by the interaction process implemented in the model, so that simulated communication patterns compare to the real--world exchange processes.

In a qualitative sense, the degree distributions of the interaction networks in the phase transition shown throughout this paper (Fig.~\ref{fig:DegreeDistribution} and \ref{fig:All.N1000to10000}) relate to observations made on real--world communication networks.
A scaling according to a power law as shown for call graphs by Onnela et al. in \cite{Onnela2007a,Onnela2007b} (compare Fig. 1A in~\cite{Onnela2007b}) as well as by Aiello et al. in \cite{Aiello2000} (Fig. 1 and 2), is observed in the present model of opinion exchange.
Another confirming observation is the percentage of nodes in the largest connected component.
In the mobile network study (\cite{Onnela2007a,Onnela2007b}) it consists of 84\% of the nodes and in our networks this percentage is found to be in between 75 and 85\%.
%Though this is still far from proving the validity of the model, it is a first indication of it.
The qualitative similarity may be a useful starting point for future calibration of the opinion model.

One has to take into account in such a comparison that the real--world examples consist of several millions of nodes (e.g., 3.9 million in \cite{Onnela2007a,Onnela2007b}) and that the largest network created in the course of this study has (only) 10000 participants.
However, as the previous section revealed, the degree scaling did not change with an increasing population size.

The emergence of very complex, non--trivial social structures from simple opinion exchange processes becomes also visible in the example network presented at the very end of this paper (in Fig~\ref{fig:Network1000}).
Agents do not have any knowledge of the global properties of the network nor is there any routine by which they urge to improve their position in the network.
Nevertheless, we observe the formation of various clusters that among themselves are strongly connected, and also the emergence of individuals that connect between different clusters becomes visible.
Having a high centrality, but only a small number of social contacts to sustain, such nodes are considered to have a high importance in the social network (compare~\cite{Gross2008} and references therein).

While the generation of networks provides us with the possibility of linking the simulation model and reality at the level of global structure, the use of bit--strings in the opinion representation allows for a connection of agent opinions and real survey data.
For instance, questions like >>are you currently satisfied with the work of politician X?<< or >>which of the following issues you think should be regulated/not regulated by the government?<< can be used to set up an agent population with opinions distributed according to the results of a questionnaire.
Repeated surveys might even allow for comparison of real data to the simulated opinion exchange dynamics.

\section*{Future Developments}

Besides being scale--free, real--world networks are typically characterized by some other statistical properties.
Many posses a relatively small average diameter (small--world property) and a high clustering.
So far our analysis did not consider these properties in the analysis of the emerging network of opinion exchange activity.
Classifying the networks using additional statistical network measures, such as betweenness centrality, diameter and clustering measures like the continuous clustering coefficient introduced in~\cite{Vilela2003}, is one essential issue to be addressed in the future.

\begin{figure}[ht]
    \centering
        \includegraphics[width=0.90\linewidth]{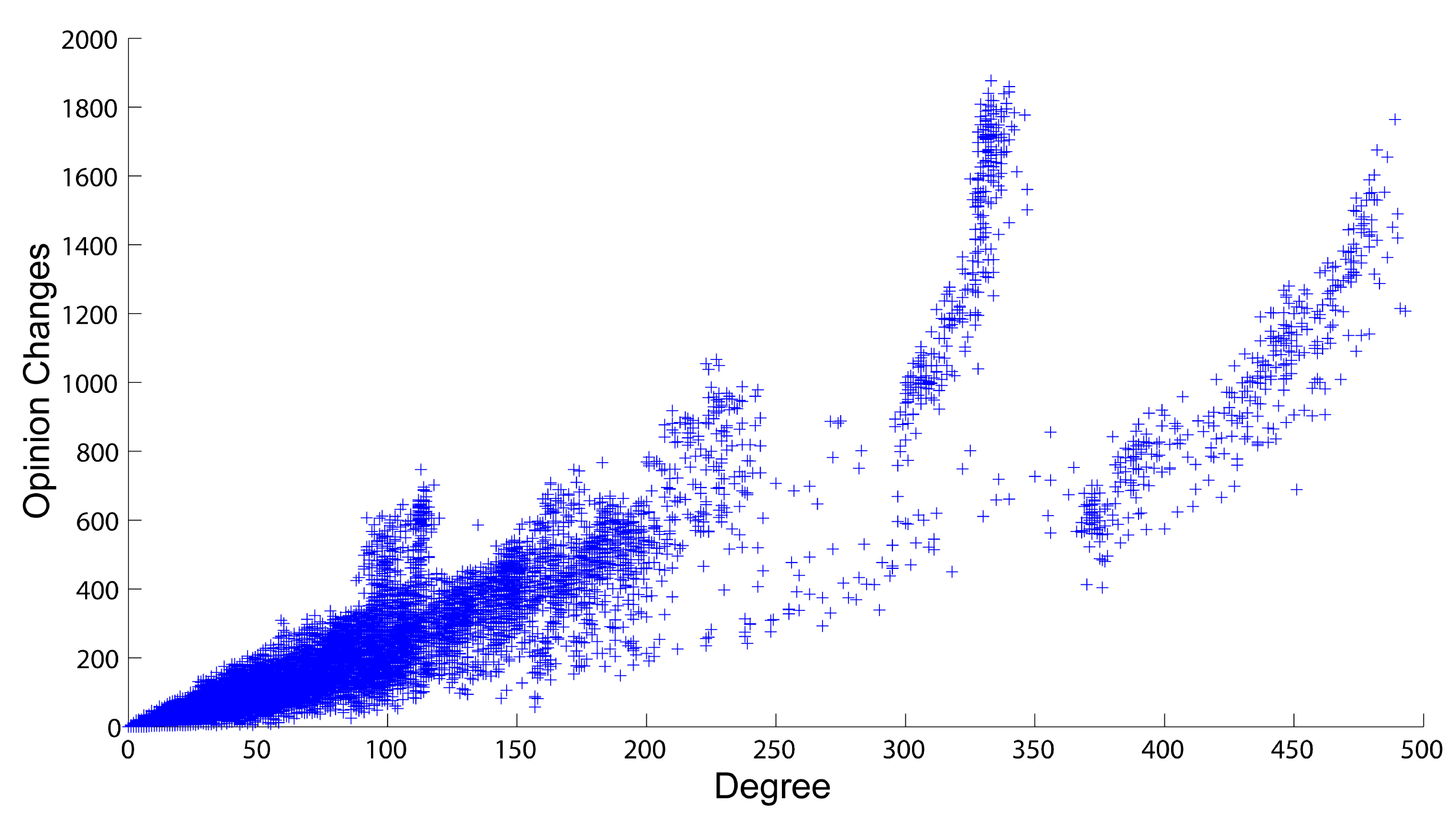}
    \vspace{-11pt}
    \caption{Number of opinion changes to the degree.}
    \label{fig:DegToChanges}
\end{figure}

\begin{figure}[ht]
    \centering
            \vspace{-11pt}
        \includegraphics[width=0.8\textwidth]{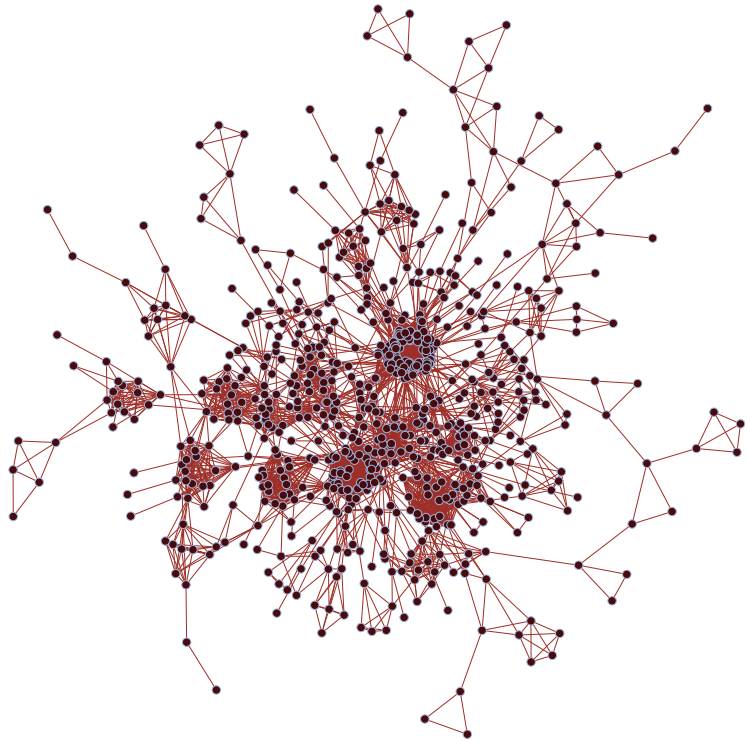}
        \vspace{-44pt}
    \caption{A network of communication formed by the interaction of 1000 agents ($k=20$ and $d_I = 3$). We can observe the formation of various clusters that among themselves are strongly connected. We also observe the emergence of individuals that connect between clusters which gives them a high importance in the social network. All in all, this shows that very complex social structures can emerge from opinion exchange processes and individual communication. Image produced with the Network Workbench Tool~\cite{NWB2006}.}
    \label{fig:Network1000}
\vspace{-33pt}
\end{figure}

Looking at opinion dynamics from the perspective of networks allows for a series of additional analyses, one of which is shown in Fig.~\ref{fig:DegToChanges}.
The data stored during the 100 simulation runs for $N = 1000$ and $d_I = 3$ is used in this analysis.
In Fig.~\ref{fig:DegToChanges}, the number of opinion changes is plot with respect to the degree of the nodes.
Surprisingly, we observe strong irregularities in the exchange behaviour.
For node degrees around 100 and around 320 significantly more opinion exchange is observed than for neighbouring degrees.
This might be an indication that critical connectivity level exists, and that the nodes once they reached this connectivity enter a larger community of agents which gives them a whole group of new communication partners.
Such a reasoning may also explain the scaling of the degree distribution for degrees larger than 100.
Future research has to clarify these effects.

\enlargethispage{\baselineskip}
Due to the high flexibility of the bit--string description adopted in this work, a series of model refinements can be implemented without much effort.
For instance, splitting the string into two and using one sub--string for (say) the private live and the other for the professional relations will allow to generate two different networks and to study the interrelation between the two.
Considering parts of the string as fixed, accounting for fixed attributes like gender, can also be a reasonable extension to the model.
And along this line, a recent study showed that cross--gender communication is more frequent and more intensive than communication among the same gender~\cite{Leskovec2008}, which reveals that besides similarity also differences attract in some cases.
An adaptation of the interaction rule such that for a part of the bit--string difference is more appealing than similarity is another candidate for future refinement of the model.

Another future development will concern the validation of the opinion dynamics model. 
Initial assumptions of the model can be grounded on previous research and on literature, and they can also be validated through some empirical studies. 
In this case, the assumptions of similarity and imitation follow Axelrod's well--known and discussed dissemination of culture model. 
The model outcomes which are the communication network and the group sizes will be the subject of future research, aiming to compare them to some real--world networks and group configurations as outlined in the previous section.

\section*{Conclusions}

Like other authors in the opinion formation field, we explore the mechanisms through which similarity leads to interaction and interaction leads to still more similarity.
Network properties seem a natural way to describe the structural patterns that come out from those multi--agent interactions. 
In this context, the main purpose of this paper is to discuss the potential of networks to emerge endogenously from local interactions without explicitly specifying rules for network linking.
This facilitates comparisons of model results to real--world social networks.

We show that complex network structures emerge from a simple process of communication between individuals that have no information on the global properties of the network.
This indicates that a crucial role in the formation of social structures and associations is played by the mutual interactions among individuals with diverse opinions, attitudes and lifestyles.

In the context of opinion dynamics, considering which networks result from simulations with different model parameters made visible
that the phase transition from highly fragmented public opinion to homogeneity in models of social influence is due to the
communication activity that is allowed by a certain parameter constellation. A few critical values play the fundamental role.

We have followed a biological inspiration, where opinion change compares to a mutation mechanism that allow for the adoption of a new position with respect to a certain issue in the agents mind.
Such an abstract bit--string approach has been used in the context of labour market analysis~\cite{Araujo2008} where bit--strings represent job offers and worker skills.
It was also applied to model innovation in a market--oriented context where producers and consumers try to improve their matching in what concerns products and needs~\cite{Araujo2009}.
We envision that the application of the notion of opinion exchange presented in this paper could bring relevant improvement to the way the underlying population structures were represented in those earlier approaches.

\section*{Acknowledgements}

The authors kindly acknowledge the helpful comments of two anonymous reviewers,
making us aware of the importance to study different population sizes.
This project was supported by FCT Portugal, Project PDCT/EGE/60193/2004.

%\bibliographystyle{ws-acs}
%\bibliography{OFaCN_BIB}

\begin{thebibliography}{10}
\providecommand{\urlprefix}{}
\expandafter\ifx\csname urlstyle\endcsname\relax
  \providecommand{\doi}[1]{doi:\discretionary{}{}{}#1}\else
  \providecommand{\doi}{doi:\discretionary{}{}{}\begingroup
  \urlstyle{rm}\Url}\fi

\bibitem{Aiello2000}
Aiello, W., Chung, F. R.~K., and Lu, L., A random graph model for massive
  graphs, in \emph{STOC} (2000), pp. 171--180.

\bibitem{Albert2002}
Albert, R. and Barabasi, A.-L., Statistical mechanics of complex networks,
  \emph{Reviews of Modern Physics} \textbf{74} (2002) 47--97.

\bibitem{Amblard2004}
Amblard, F. and Deffuant, G., The role of network topology on extremism
  propagation with the relative agreement opinion dynamics, \emph{Physica A:
  Statistical Mechanics and its Applications} \textbf{343} (2004) 725--738.

\bibitem{Amblard2008a}
Amblard, F. and Jager, W., Network shapes resulting from different processes of
  interaction, in \emph{proceedings of The 5th Conference of the European
  Social Simulation Association, Italy} (2008).

\bibitem{Araujo2008}
Araújo, T. and Weisbuch, G., The labour market on the hypercube, \emph{Physica
  A, Statistical Mechanics and its Applications} \textbf{387} (2008)
  1301--1310.

\bibitem{Araujo2009}
Ara\'{u}jo, T. and Mendes, V.~R., Innovation and self-organization in a
  multiagent model, \emph{Advances in Complex Systems} \textbf{12} (2009)
  233--253.

\bibitem{Axelrod1997}
Axelrod, R., The dissemination of culture: A model with local convergence and
  global polarization, \emph{The Journal of Conflict Resolution} \textbf{41}
  (1997) 203--226.

\bibitem{Barrat2004}
Barrat, A., Barth\'{e}lemy, M., Pastor-Satorras, R., and Vespignani, A., The
  architecture of complex weighted networks, \emph{Proceedings of the National
  Academy of Sciences of the United States of America} \textbf{101} (2004)
  3747--3752.

\bibitem{Castellano2009}
Castellano, C., Fortunato, S., and Loreto, V., Statistical physics of social
  dynamics, \emph{Reviews of Modern Physics} \textbf{81} (2009) 591--646.

\bibitem{Castellano2000}
Castellano, C., Marsili, M., and Vespignani, A., Nonequilibrium phase
  transition in a model for social influence, \emph{Physical Review Letters}
  \textbf{85} (2000) 3536--3539.

\bibitem{Chung2006}
Chung, F. and Lu, L., \emph{Complex Graphs and Networks} (American Mathematical
  Society, 2006).

\bibitem{Deffuant2000}
Deffuant, G., Neau, D., Amblard, F., and Weisbuch, G., Mixing beliefs among
  interacting agents, \emph{Advances in Complex Systems} \textbf{3} (2001)
  87--98.

\bibitem{desJardins2008}
desJardins, M., E.~Gaston, M., and Radev, D., Introduction to the special issue
  on ai and networks, \emph{AI Magazine} \textbf{29} (2008) 11--15.

\bibitem{Epstein2006}
Epstein, J.~M., \emph{Remarks on the Foundations of Agent-Based Generative
  Social Science}, chapter~34 (Elsevier B. V., 2006).

\bibitem{Festinger1954}
Festinger, L., A theory of social comparison processes, \emph{Human Relations}
  \textbf{7} (1954) 117--140.

\bibitem{Galam2004}
Galam, S., The dynamics of minority opinions in democratic debate,
  \emph{Physica A: Statistical and Theoretical Physics} \textbf{336} (2004)
  56--62.

\bibitem{Gross2008}
Gross, T. and Blasius, B., Adaptive coevolutionary networks: a review,
  \emph{Journal of The Royal Society Interface} \textbf{5} (2008) 259--271.

\bibitem{Holme2006}
Holme, P. and Newman, M., Nonequilibrium phase transition in the coevolution of
  networks and opinions, \emph{Physical Review E} \textbf{74} (2006) 1--5.

\bibitem{Jiang2007}
Jiang, L.-l., Hua, D.-y., and Chen, T., Nonequilibrium phase transitions in a
  model with social influence of inflexible units, \emph{Journal of Physics A:
  Mathematical and Theoretical} \textbf{40} (2007) 11271--11276.

\bibitem{Leskovec2008}
Leskovec, J. and Horvitz, E., Planetary-scale views on a large
  instant-messaging network, in \emph{WWW '08: Proceeding of the 17th
  international conference on World Wide Web} (ACM, New York, NY, USA, 2008),
  ISBN 978-1-60558-085-2, pp. 915--924,
  \doi{http://doi.acm.org/10.1145/1367497.1367620}.

\bibitem{NWB2006}
NWB-Team, Network workbench tool (2006),
  \urlprefix\url{http://nwb.slis.indiana.edu}, indiana University, Northeastern
  University, and University of Michigan.

\bibitem{Onnela2007b}
Onnela, J.~P., Saramaki, J., Hyvonen, J., Szabo, G., Lazer, D., Kaski, K.,
  Kertesz, J., and Barabasi, A.~L., Structure and tie strengths in mobile
  communication networks, \emph{Proceedings of the National Academy of
  Sciences} \textbf{104} (2007) 7332--7336.

\bibitem{Onnela2007a}
Onnela, J.-P., Saram{\"a}ki, J., Hyv{\"o}nen, J., Szab{\'o}, G., De~Menezes, M.~A., Kaski,
  K., Barab{\'a}si, A.-L., and Kert{\'e}sz, J., Analysis of a large-scale weighted
  network of one-to-one human communication, \emph{New Journal of Physics}
  \textbf{9} (2007) 1--27.

\bibitem{Vilela2003}
R.~Vilela~Mendes, T.~A. and {a}, F.~L., Reconstructing an economic space from a
  market metric, \emph{Physica A} \textbf{323} (2003) 635--50.

\bibitem{Rosvall2007}
Rosvall, M. and Sneppen, K., Dynamics of opinions and social structures,
  http://arxiv.org/abs/0708.0368 (2007).

\bibitem{Schweitzer2003}
Schweitzer, F., Meinungsbildung, kommunikation und kooperation aus
  physikalischer perspektive, \emph{Physik Journal} \textbf{2} (2003) 57--62.

\bibitem{Schweitzer2008}
Schweitzer, F. and Behera, L., Nonlinear voter models: The transition from
  invasion to coexistence, \emph{The European Physical Journal B - Condensed
  Matter and Complex Systems} \textbf{67} (2008) 301--318.

\bibitem{Sobkowicz2009}
Sobkowicz, P., Modelling opinion formation with physics tools: Call for closer
  link with reality, \emph{Journal of Artificial Societies and Social
  Simulation} \textbf{12} (2009) 1.

\bibitem{SznajdWeron2004}
Sznajd-Weron, K., Dynamical model of ising spins, \emph{Phys. Rev. E}
  \textbf{70} (2004) 037104.

\bibitem{Zanette2006}
Zanette, D. and Gil, S., Opinion spreading and agent segregation on evolving
  networks, \emph{Physica D: Nonlinear Phenomena} \textbf{224} (2006) 156--165.

\end{thebibliography}

\end{document}